\documentclass[twocolumn,showpacs,preprintnumbers,amsmath,amssymb,floatfix]{revtex4}
\usepackage{graphicx}

\newcommand{\qnm}{\text{\tiny QNM}}

\begin{document}

\title{Quasinormal Modes, the Area Spectrum, and Black Hole Entropy}
\date{\today}
    \author{Olaf Dreyer}
    \email{odreyer@perimeterinstitute.ca}
    \affiliation{Perimeter Institute for Theoretical Physics, 35 King
    Street North, Waterloo, Ontario N2J 2W9, Canada}

\begin{abstract} The results of canonical quantum gravity
concerning geometric operators and black hole entropy are beset by
an ambiguity labelled by the Immirzi parameter. We use a result
from classical gravity concerning the quasinormal mode spectrum of
a black hole to fix this parameter in a new way. As a result we
arrive at the Bekenstein - Hawking expression of $A/4 l_P^2$ for
the entropy of a black hole and in addition see an indication that
the appropriate gauge group of quantum gravity is SO(3) and not
its covering group SU(2).
\end{abstract}


\maketitle

\subsection{Introduction}
The canonical approach to quantum gravity (see
\cite{Thiemann:2001yy} for a recent review) boasts two remarkable
sets of results. It has quantum operators for area and volume that
have discrete spectra and it puts forward a derivation of the
entropy of a black hole. Both these results are plagued by the
existence of one free parameter. Up to now the only way to fix
this ambiguity was to use the result for the black hole entropy
and adjust it to the Bekenstein - Hawking result. This then also
fixes the ambiguity in the spectra of the geometrical operators.
In this note we will put forward an independent argument to fix
the ambiguity rendering the black hole calculation a true
prediction of the theory.

A basis for the Hilbert space of canonical gravity is given by
spin networks. These are graphs whose edges are labelled by
representations of the gauge group of the theory. In the case of
gravity this group is taken to be SU(2) and the representations
are thus labelled by positive half-integers $j = 0, 1/2, 1, 3/2,
\ldots$. If a surface is intersected by an edge of such a spin
network carrying the label $j$ the surface acquires the area
\cite{rovsmo, ashlew}
\begin{equation}\label{eqn:area}
  A(j) = 8\pi l_P^2 \gamma \sqrt{j(j+1)},
\end{equation}
where $l_P$ is the Planck length and $\gamma$ is the so called
Immirzi parameter \cite{Immirzi:1996dr}. This is the ambiguity we
spoke of before. It parameterizes an ambiguity in the choice of
canonically conjugate variables that are to be quantized. There is
no a priori reason to fix this parameter to any particular value.

\begin{figure}
  \begin{center}
  \includegraphics[height=3.5cm]{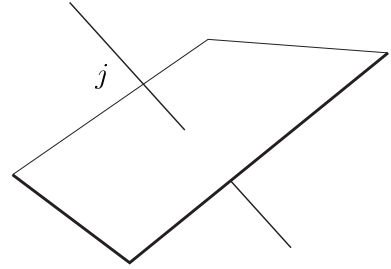}
  \caption{\small{In canonical quantum gravity the area of a surface is quantized.%
  If a surface intersects a spin network edge with label%
  $j$ it acquires an area of $8\pi\gamma l_P^2\sqrt{j(j+1)}$. The parameter%
  $\gamma$ is called the Immirzi parameter.}}\label{fig:areaj}
  \end{center}
\end{figure}

The only argument so far that could be used to fix this parameter
comes from black hole entropy. Given a black hole horizon one can
think of the area of the horizon as being a consequence of a large
number of spin network edges puncturing the surface. (For a
detailed account of black hole entropy in canonical quantum
gravity see \cite{smolin, rovelli, krasnov, abck, abk}. The review
\cite{Thiemann:2001yy} and the citations contained therein are
also helpful.) Each edge with spin $j$ contributes the amount of
area given by formula (\ref{eqn:area}) to the whole area. On the
horizon such a puncture also increases the dimensionality of the
Hilbert space of the theory living on the boundary. Each puncture
of an edge with spin $j$ increases the dimension by a factor of
$2j +1$, i.e. by the dimension of the spin $j$ representation. If
there there is a large number $N$ of edges with spins $j_i$, $i=1,
\ldots, N$, intersecting the horizon the dimension of the boundary
Hilbert space is
\begin{equation}
  \prod_{i=1}^N (2j_i + 1).
\end{equation}
The entropy of a black hole with a given area $A$ is then given by
the logarithm of the dimension of the Hilbert space of the
boundary theory. It can be shown that the statistically most
important contribution comes from those configurations in which
the lowest possible spin dominates. Let us denote this spin by
$j_{\min}$. The entropy is then
\begin{equation}\label{eqn:entropy}
    S = N \ln (2j_{\min}+1),
\end{equation}
where $N$ can be calculated from the area $A$ of the black hole
and from the amount of area $A(j_{\min})$ contributed by every
puncture. One obtains
\begin{equation}
  N = \frac{A}{8\pi l_P^2\gamma\sqrt{j_{\min}(j_{\min}+1)}}.
\end{equation}
Equating the result of equation (\ref{eqn:entropy}) with the known
Bekenstein - Hawking result then gives a value for the Immirzi
parameter. The lowest non-trivial representation of SU(2) has spin
$j_{\min}=1/2$ and one obtains the value $\ln 2/\pi \sqrt{3}$ for
the Immirzi parameter.

In this note we want to fix the Immirzi parameter in a way that is
independent from the black hole considerations outlined above. For
this we will make use of an observation by Hod \cite{hod} on the
quasinormal ringing modes of a black hole. Aside from arriving at
a new value for the Immirzi parameter we will argue that the
lowest admissible spin should be $j_{\min}=1$ and not
$j_{\min}=1/2$.

\subsection{Quasinormal Modes}
\begin{figure}
  \begin{center}
  \includegraphics[height=5.5cm]{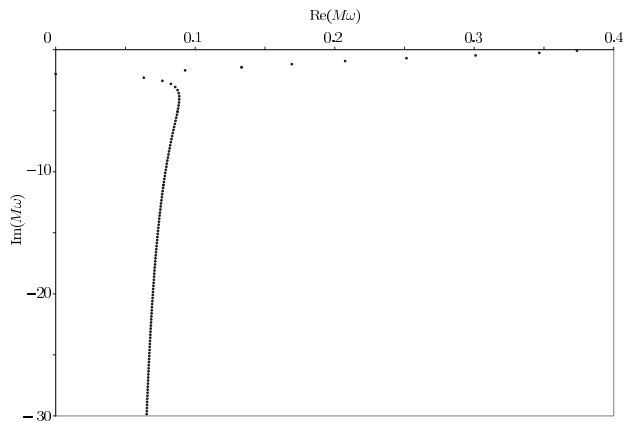}
  \caption{\small{This figure shows the first 124 quasinormal
  mode frequencies of a Schwarzschild black hole.}}\label{fig:qnms}
  \end{center}
\end{figure}

The reaction of a back hole to a perturbation will be dominated by
a set of damped oscillations called quasinormal modes. They appear
as solutions to the perturbation equation of the Schwarzschild
geometry found by Regge and Wheeler, and Zerilli. Figure
\ref{fig:qnms} shows the first quasinormal mode frequencies in the
complex $\omega$ plane. (See the review articles by Nollert
\cite{nollertrev}, and Kokkotas and Schmidt \cite{kokkotas} for
more information about quasinormal modes.)

For large damping, i.e. for a large imaginary part of $\omega$,
the real part of the frequency approaches a nonzero value and the
imaginary part becomes equally spaced. Nollert \cite{nollert}
found the following limiting behavior of the quasinormal mode
frequencies:
\begin{equation}
  M \omega = 0.04371235 + \frac{i}{4}\left( n +
  \frac{1}{2}\right),
\end{equation}
where $M$ is the mass of the black hole. This relation was later
confirmed by Andersson \cite{andersson}. Crucial for our argument
is an observation made by Hod \cite{hod}. He remarked that the
constant real part of the quasinormal frequencies is equal to
\begin{equation}
  \frac{\ln 3}{8\pi}.
\end{equation}
So far the evidence for this relation is purely numerical. What is
remarkable here is the appearance of the $\ln 3$ term. Since we
will use this frequency later on, we denote it by $\omega_\qnm$:
\begin{equation}\label{eqn:omqnm}
    \omega_\qnm = \frac{\ln 3}{8 \pi M}
\end{equation}

\subsection{The Area Spectrum}
Using $\omega_\qnm$ we will now fix the ambiguity of the area
spectrum. According to Bohr's correspondence principle an
oscillatory frequency of a classical system should be equal to a
transition frequency of the corresponding quantum system. The most
natural candidate for a transition of the quantum black hole as
described above is the appearance or disappearance of a puncture
with spin $j_{\min}$. The area of the black hole would then change
by an amount given by equation ({\ref{eqn:area}) where $j =
j_{\min}$:
\begin{equation}\label{eqn:deltaa}
    \Delta A = A(j_{\min}) = 8 \pi l_P^2 \gamma \sqrt{j_{\min}(j_{\min}+1)}
\end{equation}

We can now fix the Immirzi parameter $\gamma$ appearing in this
equation by requiring that the change $\Delta M$ in the mass
corresponding to this change in area equals the energy of a
quantum with frequency $\omega_\qnm$, i.e. we will fix $\gamma$ by
setting
\begin{equation}\label{eqn:equal}
    \Delta M = \hbar \omega_\qnm = \frac{\hbar\ln 3}{8 \pi M}.
\end{equation}
Since the area $A$ and the mass $M$ of a Schwarzschild black hole
are related by
\begin{equation}\label{eqn:aandm}
    A = 16 \pi M^2
\end{equation}
the mass change of equation (\ref{eqn:equal}) translates into the
area change
\begin{equation}\label{eqn:areaj}
  \Delta A = 4\ln3\; l_P^2.
\end{equation}
Comparing this with equation (\ref{eqn:deltaa}) gives the desired
expression for the Immirzi parameter $\gamma$:
\begin{equation}\label{eqn:newgam}
  \gamma = \frac{\ln 3}{2\pi\sqrt{j_{\min}(j_{\min}+1)}}
\end{equation}
In the next section we will go further and also fix $j_{\min}$.

\subsection{The Entropy}
The entropy is given by the logarithm of the dimension of the
Hilbert space of the boundary theory. As before we have
\begin{equation}\label{eqn:sagain}
  S = N \ln (2j_{\min} +1),
\end{equation}
where $N$ is the number of punctures given by
\begin{equation}\label{eqn:n}
  N = \frac{A}{A(j_{\min})}.
\end{equation}
In the previous section we argued that $\Delta A = A(j_{\min}) =
4\ln3\; l_P^2$ and thus
\begin{equation}\label{eqn:sfinal}
  S = \frac{A}{4 l_P^2}\frac{\ln (2j_{\min} +1)}{\ln 3}.
\end{equation}
A complete agreement with the Bekenstein - Hawking result is
obtained if we set
\begin{equation}\label{eqn:setjmin}
  j_{\min} = 1.
\end{equation}
This then also fixes the value for the Immirzi parameter:
\begin{equation}\label{eqn:newgamma}
  \gamma = \frac{\ln 3}{2\pi \sqrt{2}}
\end{equation}

\subsection{Discussion}
The quasinormal mode spectrum of a black hole singles out the
frequency $\omega_\qnm$ of equation (\ref{eqn:omqnm}). For large
damping $\omega_{\qnm}$ is the real part of a quasinormal mode
frequency. This uniqueness of $\omega_\qnm$ strongly suggests that
this frequency should also play a role in the quantum theory of a
black hole. Invoking Bohr's correspondence principle we argued
that $\omega_\qnm$ should appear as a transition frequency in the
quantum theory.

In canonical quantum gravity the area of a macroscopically large
black hole arises through a large number of intersection of the
black horizon surface with spin network edges with the same low
spin $j_{\min}$. The natural candidate for a quantum transition is
thus the appearance or disappearance of a puncture of spin
$j_{\min}$. By equating the mass change corresponding to this area
change with the energy $\hbar \omega_\qnm$ of a quantum with
frequency $\omega_\qnm$ we can fix the Immirzi parameter as a
function of the lowest spin $j_{\min}$.

Using these results we were then able to calculate the entropy of
a black hole. This entropy agrees with the Bekenstein - Hawking
result provided one chooses
\begin{equation}
  j_{\min} = 1. \nonumber
\end{equation}

We want to point out how remarkable this result is. The purely
classical frequency $\omega_\qnm$ contains exactly the factor of
$\ln 3$ required to cancel the same factor appearing in the
formula for the entropy. The classical theory and the quantum
theory of a black hole conspire here to give both the Bekenstein -
Hawking entropy and a prediction about the lowest admissible spin
$j_{\min}$.

Next we want to discuss the significance of the result $j_{\min} =
1$. There are two possible explanations. The first explanation can
be that the gauge group to consider is SO(3) and not SU(2). Since
the unitary representations of SO(3) are labelled by integers the
lowest allowed spin would be $j_{\min} = 1$. This explanation is
perfectly plausible since the use of SU(2) connections is not
motivated by physical consideration but rather by convenience. In
most of the calculations the gauge group appears only through its
Lie algebra and it is then not surprising that the representations
of this algebra are used. In the case of canonical quantum gravity
this means that it is not SO(3) that is used but its simply
connected covering group SU(2).

Another explanation can be that there is a physical reason why
spin $1/2$ punctures should not be counted. No such reason is
known so far.

We pointed out above that the area of a macroscopically large
black hole is mainly due to a large number of punctures with one
low spin $j_{\min}$. This means that the possible transitions are
dominated by the appearance and disappearance of one puncture with
spin $j_{\min}$. We had to invoke this statistical argument here
since the area spectrum obtained from (\ref{eqn:area}) is nearly
continuous for a large area. This is because the possible values
for the area coming from one puncture depend on the spin in an
irrational manner through a square root. The possible values for
the area obtained by a large number of punctures then become dense
for large areas.

One can argue that this means that the quantization of the area
operator has to be changed. Recently such quantizations have been
proposed \cite{alekseev}. The spectrum obtained is of the form
\begin{equation}\label{eqn:arealt}
  A = \tilde\gamma l_P^2 (j + 1/2).
\end{equation}
Since this spectrum is equally spaced the possible values for the
area of a macroscopically large black hole are also equally spaced
with the same spacing. Using the same arguments as above one can
then fix the constant $\tilde\gamma$ to be equal to $4\ln 3$.

The problem with this approach is that it does not give the
Bekenstein - Hawking entropy if one follows the same procedure as
above. Since the difference $\Delta A$ between two adjacent area
values does not equal the area $A(j_{\min})$ the entropy is not
equal to $A/4 l_P^2$. With $j_{\min}$ equal to unity one obtains
an entropy of $A/6 l_P^2$. This problem arises because of the
$1/2$ in formula (\ref{eqn:arealt}). If one would remove this
constant one would again have $\Delta A = A(j_{\min})$ and one
would again obtain the Bekenstein - Hawking result.

For the argument in this paper to work we need
\begin{equation}\label{ean:daamin}
  \Delta A = A(j_{\min}),
\end{equation}
where $\Delta A$ is equal to $4\ln 3\; l_P^2$. This was obtained
from the quasinormal mode considerations.

The treatment of black holes in canonical quantum gravity is so
far confined to non-rotating black holes. The inclusion of
rotation has proven to be a non-trivial task. Maybe the insight
gained here by looking at quasinormal modes can shed some light on
the problem. The quasinormal mode spectrum of a rotating black
hole is considerably more complicated than that of a non-rotating
black hole (see the reviews \cite{nollertrev, kokkotas} and also
the work by Leaver \cite{leaver}). If a connection between
classical and quantum theory also holds true for rotating black
holes, a more complicated set of quantum transitions is required.

When the lowest allowed spin is $1/2$ there is a nice similarity
between the picture coming from canonical quantum gravity and
Wheeler's  ``it from bit" picture \cite{wheeler}. It seems that we
are led to a somewhat more complicated picture in which the
degrees of freedom are not just $\pm 1$ as in Wheeler but $0, \pm
1$.

Finally we want to point out that the result we use in all our
derivations, namely that $\omega_\qnm = \ln 3/8\pi M $, is so far
known only through numerical calculations. The argument would be
even stronger with an analytic proof of this result.

The author would like to thank G.~Kunstatter, E.~R.~Livine,
D.~Oriti, and the members of the Perimeter Institute for fruitful
discussions.

\end{document}